 \documentclass[final,3p,times]{elsarticle}

\usepackage{graphicx,url}

\usepackage{amssymb}

\biboptions{sort&compress} 

\newcounter{bla}

\newcommand{\be}{\begin{equation}}
\newcommand{\ee}{\end{equation}}
\newcommand{\ba}{\begin{eqnarray}}
\newcommand{\ea}{\end{eqnarray}}

\def\lsim{\raise0.3ex\hbox{$\;<$\kern-0.75em\raise-1.1ex\hbox{$\sim\;$}}}
\def\gsim{\raise0.3ex\hbox{$\;>$\kern-0.75em\raise-1.1ex\hbox{$\sim\;$}}}
\def\eps{\varepsilon}
\def\theta{\vartheta}
\def\ap{\approx}

\journal{Computer Physics Communications}

\begin{document}

\begin{frontmatter}



\title{{\tt ELMAG}: A Monte Carlo simulation of electromagnetic cascades
 on the extragalactic background light and in
 magnetic fields}


\author[a]{M.~Kachelrie\ss
}
\author[a,b]{S.~Ostapchenko}

\author[c]{and R.~Tom\`as}

\address[a]{Institutt for fysikk, NTNU, Trondheim, Norway}
\address[b]{D.~V.~Skobeltsyn Institute of Nuclear Physics,
 Moscow State University, Russia}
\address[c]{II. Institut f\"ur Theoretische Physik,
   Universit\"at Hamburg, Germany}

\begin{abstract}
A Monte Carlo program for the simulation of electromagnetic cascades 
initiated by high-energy photons and electrons interacting with
extragalactic background light (EBL) is presented.  Pair production and 
inverse Compton scattering on EBL photons as well as synchrotron losses
and deflections of the charged component in extragalactic magnetic fields
(EGMF) are included in the simulation. Weighted sampling of the 
cascade  development is applied to reduce the number of secondary particles
and to speed up computations.
 As final result, the simulation procedure provides
 the energy, the observation angle, and the time delay of 
secondary cascade particles at the present epoch.
Possible applications are the study of TeV blazars and the influence of
the EGMF on their spectra or the calculation of the contribution 
from ultrahigh energy cosmic rays or dark matter to the diffuse 
extragalactic gamma-ray background. As an illustration,
we present results for deflections and time-delays relevant for
the derivation of limits on the EGMF.
\end{abstract}

\begin{keyword}
Electromagnetic cascades \sep extragalactic background light 
\sep extragalactic magnetic fields.
\end{keyword}

\end{frontmatter}



{\bf PROGRAM SUMMARY}

\begin{small}
\noindent
{\em Manuscript Title:}  
{\tt ELMAG}: A Monte Carlo simulation of electromagnetic cascades
 on the extragalactic background light and in
 magnetic fields\\
 {\em Program Title:} {\tt ELMAG 1.01} 
\\
{\em Journal Reference:}                                      \\
{\em Catalogue identifier:}                                   \\
{\em Licensing provisions:}                                   \\
{\em Programming language:}  Fortran 95                      \\
{\em Computer:}                                               
Any computer with Fortran 95 compiler                  \\ 
{\em Operating system:}  Any system with Fortran 95 compiler                \\
{\em RAM:} 4 Mbytes                                              \\
{\em Number of processors used:} arbitrary using the MPI version  \\
{\em Supplementary material:}                                
see \url{http://elmag.sourceforge.net/}
 \\
{\em Keywords:} Electromagnetic cascades, extragalactic background light, extragalactic magnetic fields  \\
{\em Classification:}        
11.3  Cascade and Shower Simulation, 11.4 Quantum Electrodynamics
                                 \\
{\em Nature of problem:}
Calculation of secondaries produced by electromagnetic cascades on the extragalactic 
background light (EBL)
   \\
{\em Solution method:}
Monte Carlo simulation of pair production and inverse Compton scattering
on EBL photons; two parametrisations from Ref.~[1] can be chosen as
EBL; weighted sampling of the cascading secondaries; recording of energy, 
observation angle and  time delay of secondary particles at the present epoch.
\\
{\em Restrictions:}
Deflections and time-delays are calculated in the small-angle approximation.\\
{\em Unusual features:}\\
{\em Additional comments:}\\
{\em Running time:}
 400 seconds for $10^3$ photons injected at redshift $z=0.2$ with energy $E=100$\,TeV using one  Intel(R) Core(TM) i7 CPU with 2.8\,GHz.
\\

\end{small}

\section{Introduction}

The Universe is opaque to the propagation of  $\gamma$-rays with energies
in the TeV region and above~\cite{pair}. Such photons are absorbed 
by pair production on the extragalactic background light 
(EBL)~\cite{Stecker06,franceschini08,Primack08,kneiske10}, consisting 
mainly of infrared light and the cosmic microwave background (CMB).
As a result the photon flux at energies $E\gsim 10$\,TeV from distant 
sources as e.g.\ blazars is significantly attenuated on the way from the 
source to the Earth. High-energy photons are however not really absorbed but
initiate electromagnetic cascades in the intergalactic space, via the 
two processes
\be
 \gamma+\gamma_b\to e^++e^-
\ee
\be
 e^\pm+\gamma_b\to e^\pm+\gamma \,.
\ee
The cascade develops very fast until it reaches the pair creation threshold
at\footnote{We use natural units, $\hbar=c=k_B=1$, throughout the text.} 
$s_{\min}=4E_\gamma\eps_\gamma=4m_e^2$ with $\eps_\gamma$ as 
the characteristic energy of the background photons $\gamma_b$. 
Electrons\footnote{We call from now on electrons and positrons collectively 
electrons.} continue to scatter on EBL
photons in the Thomson regime with an interaction length of a few kpc, 
producing photons with average energy
\be  \label{thomson}
 E_\gamma = \frac{4}{3}\frac{\eps_\gamma E_e^2}{m_e^2} 
 \ap 3\:{\rm GeV} \: \left( \frac{E_e}{1{\rm TeV}} \right)^2 
\ee 
using  $\eps_\gamma= 2.7\, T_{\rm CMB} \ap 6.3\times 10^{-4}$\,eV as the typical 
energy of CMB photons.
 
The resulting shape of the energy spectrum of the diffuse photon flux 
$J_\gamma$ can be estimated analytically~\cite{bs75} for a monochromatic
background as
$$
J_\gamma(E_\gamma) = \left\{
\begin{array}{lll}
K (E_\gamma/E_{\rm x})^{-3/2}  \quad &\mbox{for}  & E_\gamma \leq E_{\rm x}\,, \\
K (E_\gamma/E_{\rm x})^{-2}  \quad &\mbox{for}  & E_{\rm x} \leq E_\gamma\leq E_{\min}\,,  \\
0  \quad &\mbox{for}  & E_\gamma>E_{\min} \,.
\end{array}
\right.
$$
Here, $E_{\min}=m_e^2/\eps_\gamma$ is the threshold energy for 
pair-production, while  $E_\gamma\leq E_{\rm x}$ is the energy region 
where the number of electrons remains constant. Since the last generation of 
$e^+e^-$ pairs produced share the initial energy equally, $E_e=E_{\min}/2$, 
this transition energy is given by 
$E_{\rm x}=4\eps_\gamma E_{e}^2/(3m_e^2)=E_{\min}/3$.
Thus for a monochromatic background the plateau region characterised by
an $1/E_\gamma^2$ spectrum extends only over one third of an energy decade.

A better analytical description of the cascade development in the EBL uses a 
dichromatic photon gas, with $\eps_{\rm CMB}= 6.3\times 10^{-4}$\,eV and 
$\eps_{\rm IR}= 1$\,eV as typical energies for the CMB and the (second) peak 
of the IR background, respectively. Below one half of the threshold energy 
of pair production on the IR, 
$E_e\ap E_{\min, \rm IR}/2 =m_e^2/(2\eps_{\rm IR}) \ap 1.3 \times 10^{11}$\,eV, 
the number of electrons remains constant.  In the intermediate regime, 
$E_{\min, \rm IR}\lsim E \lsim E_{\min, \rm CMB}$, electrons are Compton 
scattering on CMB photons in the Thomson regime, while photons are still
producing $e^+e^-$ pairs on IR photons. Thus the energy $E_{\rm x}$ 
below which no additional electrons are injected in the 
cascade is given by
\be
 E_{\rm x} = \frac{4}{3} \,\frac{\eps_{\rm CMB} E_e^2}{m_e^2}
 = \frac{1}{3} \, \frac{\eps_{\rm CMB}}{\eps_{\rm IR}}
   E_{\min, \rm IR} \ap 50 \, {\rm MeV} \,.
\ee

\begin{figure}
\vskip-2.2cm
\begin{center}
\includegraphics[width=0.45\linewidth,angle=0]{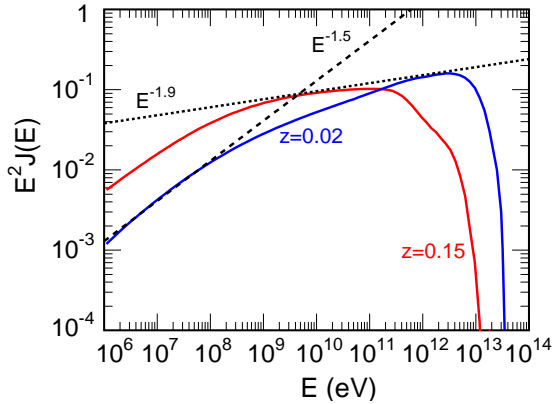}
\end{center}
\caption{
The (normalised) diffuse photon flux $E^2 J(E)$ for two sources injecting
photons with energy $E=10^{14}$\,eV  at redshift $z=0.02$ (---) and
$z=0.15$ (---), respectively.
\label{fig:dif}} 
\end{figure}

Let us now compare how well this qualitative picture agrees with the cascade 
spectrum calculated with our Monte Carlo simulation.
Figure~\ref{fig:dif} shows the results obtained with {\tt ELMAG} for the 
diffuse spectrum of secondary photons produced by photons injected 
with energy $E=10^{14}$\,eV at redshift $z=0.02$ and  $z=0.15$. 
We note first that the slope of the photon spectrum below
$E_{\rm x}$ and the obtained value of $E_{\rm x}$ agree very well
for both distances with the prediction in the simple ``dichromatic model,''
although the latter assumes an infinite number of interactions. 
In contrast, both the extension and the shape of the
plateau region are less universal: The smaller the distance to the 
source, the less pronounced occurs the steepening of the photon spectrum 
from the $E^{-1.5}$ Thomson slope towards the predicted $1/E^2$ plateau.
However,  the deviation of the slope from the prediction, $E^{-1.9}$ versus 
$E^{-2}$, is   minor already for distances of $\sim 500$\,Mpc.

An important application of electromagnetic cascades is the calculation
of various contributions to the
extragalactic diffuse gamma-ray background (EGRB). Since the Universe acts 
as a calorimeter for electromagnetic radiation, accumulating it in 
the MeV--TeV range, the measured EGRB limits all processes during the
history of the Universe
that inject electromagnetic energy above the pair creation threshold.
Examples for such processes are  photo-pion and
$p + \gamma_{\rm CMB} \to p + e^+ + e^-$ pair-production
  of UHECR protons interacting with the cosmic microwave background 
(CMB) \cite{BG88}, the decay or annihilation of (superheavy) dark matter 
or of topological defects~\cite{Kachelriess:2004ax}.
Another important application of electromagnetic cascades is the 
calculation of spectra from point sources as TeV blazars. If
the spectra of such sources extend to sufficiently high energies,
emitted photons interact with the EBL.
The charged component of these cascades is deflected by extragalactic 
magnetic fields (EGMF), leading potentially to halos around point
sources~\cite{halo1,halo2,halo3,halo4}, 
to delayed echos of flaring emission~\cite{plaga}
and influences the observed energy spectrum~\cite{spec}.
A detailed modelling of the electromagnetic
cascade process is thus not only necessary to connect the observed
energy spectra of TeV sources with their intrinsic spectra, but provides 
also information about EGMFs.

The extremely small interaction length compared to typical source distances 
from hundreds of Mpc to Gpc means that a large number $n$ of interaction 
steps has to be simulated using a Monte Carlo approach. The exponential
growth of the number $N=2^n$ of secondaries aggravates the computational
load in a brute-force Monte Carlo approach.
The Monte Carlo program presented here  uses weighted sampling of the 
cascade development to reduce efficiently the number of secondary 
particles which are traced explicitly.
 For maximally weighted sampling, the number of secondaries stays
on average constant as function of interaction steps.
Synchrotron losses and deflections of the charged component in extragalactic 
magnetic fields (EGMF) are included in the simulation too. The version
presented here is restricted to the limit of small deflections. 

\section{Modelling of the cascade process}

\subsection{Interaction rate of photons and electrons}

The interaction rate $R_{\gamma}(E,z)$ of photons with energy $E$ at
redshift $z$ can be connected to the pair-production cross section 
$\sigma_{\rm pair}(s)$ and the spectral density of background photons 
$n_{\gamma}(E,z)$ as
\ba
 R_{\gamma}(E,z) & = &  
 \frac{1}{2}\int_0^\infty \!\!\! dE'\; n_{\gamma}(E',z)
 \int_{-1}^{1} d\mu \; (1-\mu)\;\sigma_{{\rm pair}}(s)\;\Theta(s-s_{\min})
 \nonumber \\
& = & 
\frac{1}{8E^2}\int_{s_{\min}}^{s_{\max}(E)}\! ds\;s\;
\sigma_{{\rm pair}}(s)\; I_{\gamma}\!\left(\frac{s}{4E},z\right) \,,
\label{eq:lambda-gamma}
\ea
where we introduced the auxiliary function
\be
 I_{\gamma}(E_{\min},z) =  
 \int_{E_{\min}}^{E_{\max}}\!\frac{dE'}{E'^2}\; n_{\gamma}(E',z)\,.
\label{eq:I-gamma}
\ee
Here we have also assumed that the EBL, as any truly diffuse background, is
isotropic.

The c.m.~energy squared in a $\gamma\gamma$ interaction is 
$s=2EE'(1-\mu)$ with $\mu=\cos\theta$, 
while the integration limits are given by the pair production threshold 
$s_{\min}=4m_e^2$ 
and $s_{\max}(E)=4EE_{\max}$ with $E_{\max}\sim 14$\,eV as the high energy 
cutoff of the EBL background.
The well-known pair-production cross section $\sigma_{\rm pair}(s)$ is 
given by
\begin{equation}
\sigma_{\rm pair}(s)=\frac{3}{4}\,\sigma_{{\rm Th}}\, \frac{m_e^2}{s}
 \left[(3-\beta^{4})\ln\frac{1+\beta}{1-\beta}-2\beta(2-\beta^{2})\right]\!,
\label{eq:sigma-pair}
\end{equation}
with $\sigma_{\rm Th}=8\pi\alpha^2/(3m_e^2)$ as Thomson cross section and 
$\beta=\sqrt{1-4m_e^2/s}$.

Electrons emit in the Thomson regime mainly soft photons, cf.\ 
Eq.~(\ref{thomson}). To speed up the simulation, we include therefore
as discrete interactions only those which produce secondary photons above an 
arbitrary energy threshold $E_{{\rm thr}}$. The remaining soft
interactions are integrated out and included as continuous energy loss.
Thus we define the interaction rate $R_{e}(E,z)$ of an electron with
energy $E$ at redshift $z$ as
\begin{eqnarray}
  R_{e}(E,z) & = &
  \frac{1}{2} \int_{0}^\infty\! dE'\: n_{\gamma}(E',z)
  \int_{-1}^{1}\! d\mu \; (1-\beta \,\mu ) \;
  \sigma_{{\rm C}}\!\left(s,\varepsilon\right)
  \Theta(s-s_{\min}(\varepsilon))
\nonumber \\
& = & \frac{1}{8\beta \,E^2}
   \int_{s_{\min}(\varepsilon)}^{s_{\max}(E)}\! ds\; (s-m_{e}^{2})\;
   \sigma_{{\rm C}}\!\left(s,\varepsilon\right)
   I_{\gamma}\!\left(\frac{s-m_{e}^{2}}{2E(1+\beta)},z\right)
\label{eq:lambda-el}
\end{eqnarray}
with $\varepsilon=E_{{\rm thr}}/E$, $\beta=\sqrt{1-m_{e}^{2}/E^{2}}$,
and $s=m_{e}^{2}+2EE'(1-\beta\mu)$. The integration limits are given 
by $s_{\min}(\varepsilon)=m_{e}^{2}/(1-\varepsilon)$ and
$s_{\max}(E)=m_{e}^{2}+2EE_{\max}(1+\beta)$, while
the Compton scattering cross section $\sigma_{{\rm C}}(s,\varepsilon)$
integrated above the threshold $\varepsilon$ is given by
\ba
 \sigma_{{\rm C}}(s,\varepsilon) & = & 
 \frac{3}{4}\,\sigma_{{\rm Th}}\, y_{\min}\,
 \frac{y_{\max}-y_{\min}}{1-y_{\min}}
 \left[\frac{\ln(y_{\max}/y_{\min})}{y_{\max}-y_{\min}}
 \left(1-\frac{4y_{\min}(1+y_{\min})}{(1-y_{\min})^{2}}\right)\right.
 \nonumber \\
 & +&  \left.\frac{4(y_{\min}/y_{\max}+y_{\min})}{(1-y_{\min})^{2}}
 +\frac{y_{\max}+y_{\min}}{2}\right] \,.
\label{eq:sig-ics}
\ea
Here, $y_{\min}=m_{e}^{2}/s$ and $y_{\max}=1-\varepsilon$ are respectively
the minimal and the maximal energy fractions of the secondary electron.

In turn, for the electron energy loss per unit distance due to the emission
of photons of energies $E<E_{{\rm thr}}$ one obtains
\ba
 \frac{dE_{\rm ICS/thr}}{dx}(s,\varepsilon) & = & 
 \frac{3}{4}\,\sigma_{{\rm Th}}\, y_{\min}\,
 \frac{1-y_{\max}}{1-y_{\min}}
 \left[\left(\frac{\ln(1/y_{\max})}{1-y_{\max}}-1\right)
 \left(1-\frac{4y_{\min}(1+2y_{\min})}{(1-y_{\min})^{2}}\right)\right.
 \nonumber \\
 & +&  \left.\frac{1}{6}(1-y_{\max})\,(1+2y_{\max})
 + \frac{2y_{\min}\,(1+2y_{\min}/y_{\max})\,(1-y_{\max})}{(1-y_{\min})^{2}}
 \right] \,.
\label{eq:eloss-ics}
\ea

\begin{figure}
\vskip-2.2cm
\includegraphics[width=0.45\linewidth,angle=0]{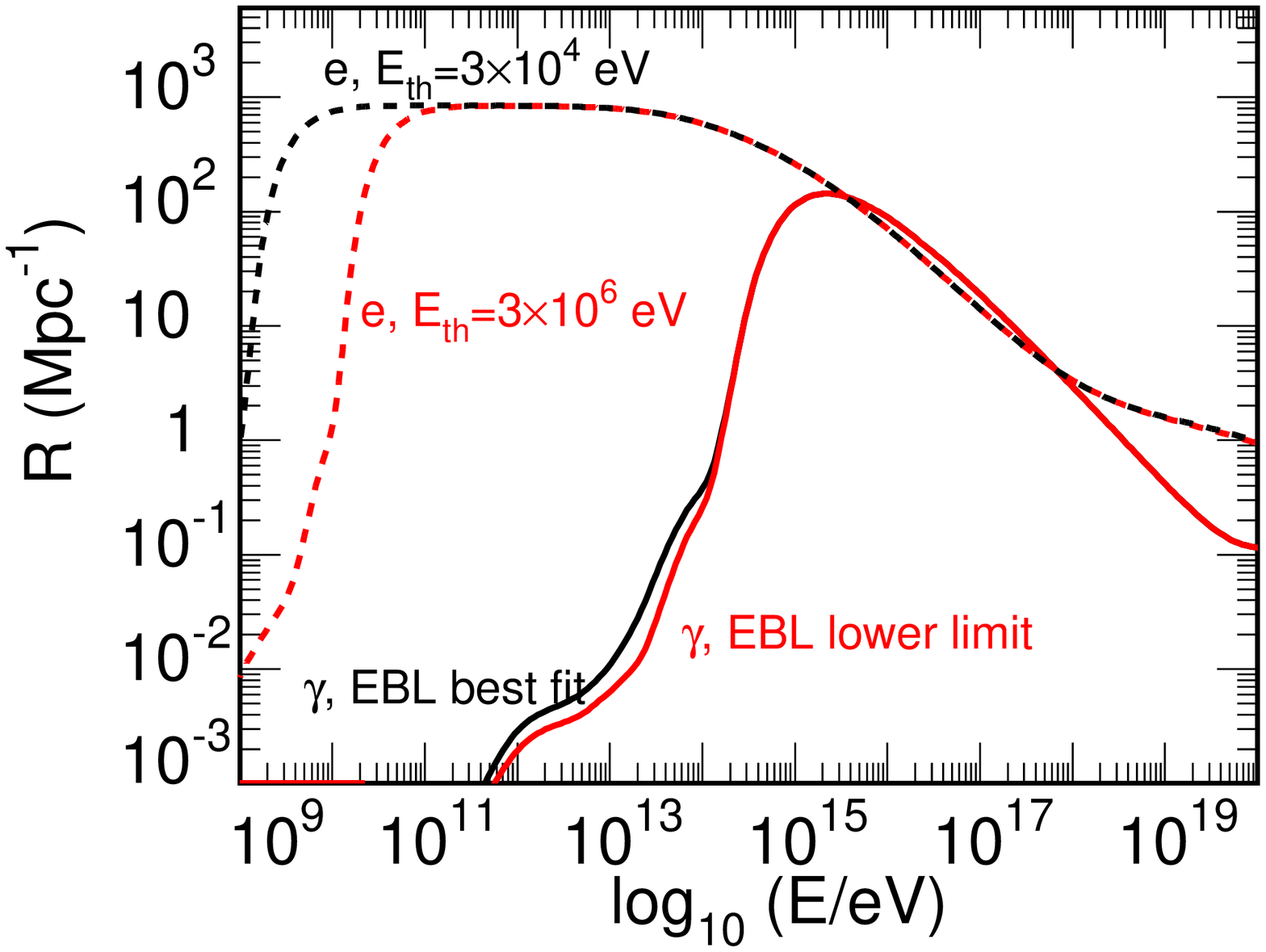}
\includegraphics[width=0.45\linewidth,angle=0]{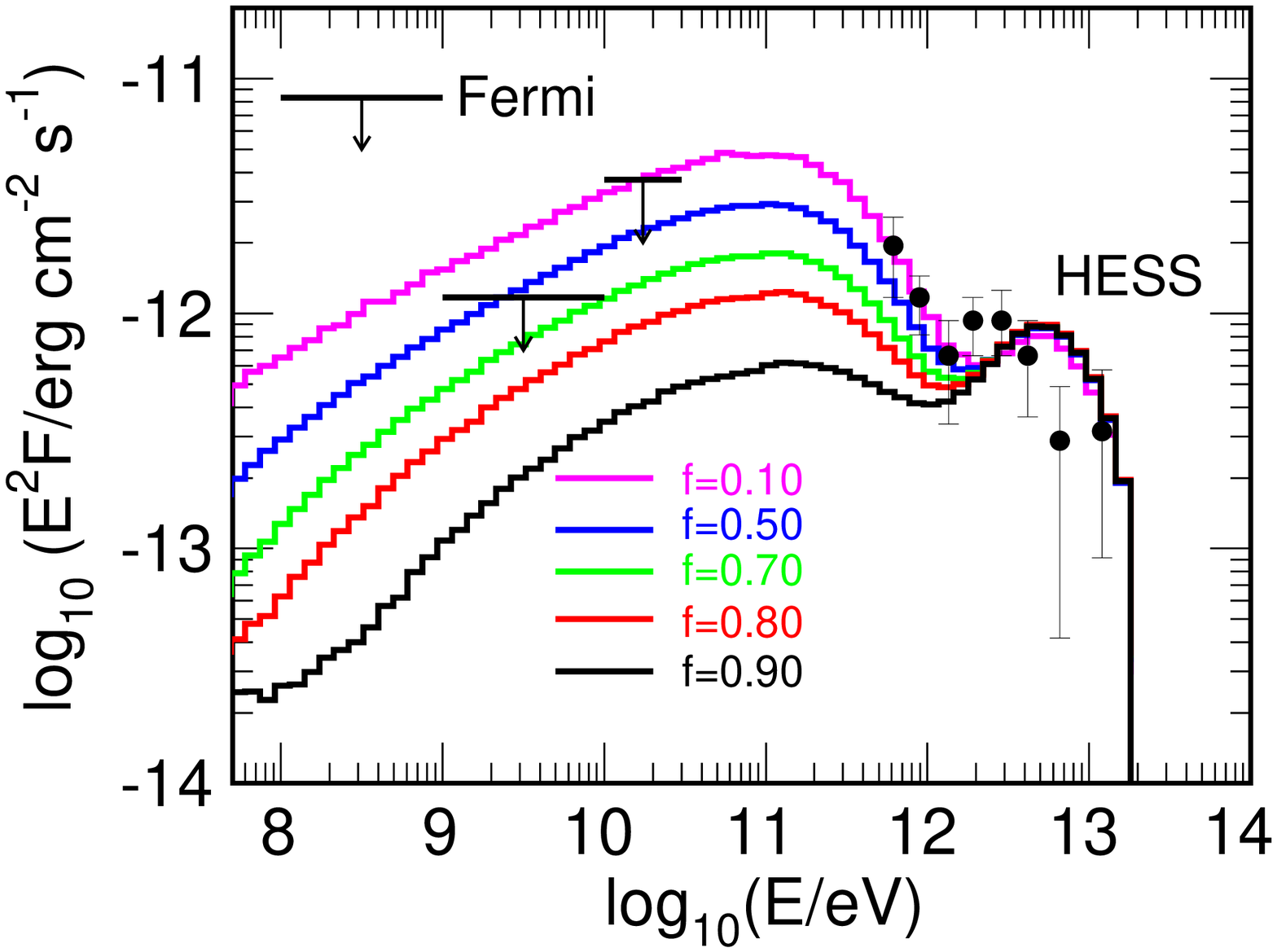}
\caption{
Left:
Interaction rate $R$ at $z=0$ as function of the energy 
$E$ for electrons with $E_{\rm th}=3\times 10^4$\,eV and 
$E_{\rm th}=3\times 10^6$\,eV and
for photons with the ``best-fit'' and the ``lower-limit'' EBL from
Ref.~\cite{kneiske10}.
Right:
Fluence contained inside the 95\% confidence contour of the 
PSF of  Fermi-LAT as function of energy for  EGMF with top-hat
profile and filling factor $f$ varying from $f=0.1$ to $f=0.9$ 
with $E_{\rm max}=20$\,TeV.
} 
\label{fig:lint}
\end{figure}

In the left panel of Fig.~\ref{fig:lint} we show the interaction rates $R_i$ 
of electrons and photons at the present epoch 
as function of energy. The difference between the ``best-fit'' and the 
``lower-limit'' EBL from Ref.~\cite{kneiske10}
becomes visible in the interaction rate $R_\gamma$ of
photons only in the energy range below $10^{14}$\,eV. 
The interaction rate $R_e$ of electrons is shown only
for the ``best-fit'' EBL but for two different values of the threshold
$E_{\rm thr}$ used in the Compton scattering cross section, 
$E_{\rm thr}=3\times 10^4$\,eV and $E_{\rm thr}=3\times 10^6$\,eV.
Note that while $E_{\rm thr}=3\times 10^6$\,eV leads already below
$10^{11}$\,eV to strong deviations from the Thomson scattering cross section,
the resulting photon spectrum is influenced by the threshold
mainly at energies in the MeV range and below, cf.~Eq.~(\ref{thomson}).

\subsection{Interactions}

The modelling of $\gamma\gamma$ and $e\gamma$ interactions starts from
sampling the c.m. energy squared $s$ of the collision according to the 
integrands of Eqs.~(\ref{eq:lambda-gamma}) and (\ref{eq:lambda-el}), 
respectively. Technically, the rejection method is used as in most other
cases to choose $s$ according to its probability distribution: The value of $s$ 
is first sampled logarithmically in the interval $[s_{\min},s_{\max}]$, then
the choice is accepted with the probability proportional to $s$ times the 
integrand of Eq.~(\ref{eq:lambda-gamma}) and (\ref{eq:lambda-el}), or 
otherwise rejected.

For given $s$, the energy fraction $y$ of the lowest energy
secondary lepton (electron or positron) in the pair production process
is sampled according to the corresponding differential cross section
\begin{eqnarray}
 \frac{d\sigma_{{\rm pair}}(s,y)}{dy} \propto  \frac{1}{y}\left[ 
 \frac{y^{2}}{1-y}+1-y+\frac{1-\beta^{2}}{1-y}-
 \frac{(1-\beta^{2})^{2}}{4y(1-y)^{2}}\right]
 \Big/ \left[1+2\beta^{2}(1-\beta^{2})\right] \,,
\label{eq:sigpair-dif}
\end{eqnarray}
with $\beta=\sqrt{1-4m_{e}^{2}/s}$. The other secondary lepton
has then the energy fraction $1-y$.
Similarly, the energy fraction $y$ of the secondary electron
in the inverse Compton process is sampled according to the differential
cross section
\begin{equation}
 \frac{d\sigma_{{\rm ICS}}(s,y)}{dy}\propto
 \frac{1}{y}\left[\frac{1+y^{2}}{2}-\frac{2y_{\min}(y-y_{\min})(1-y)}{y(1-y_{\min})^{2}}\right],
\label{eq:sig-ics-dif}
\end{equation}
with $y_{\min}=m_{e}^{2}/s$.

\subsection{Stacking and weighted sampling}

The produced secondary particles are then subject to a weighted
sampling procedure: A secondary particle carrying the  fraction
$y$ of the parent energy is discarded with the probability 
 $(1-y^{\alpha_{{\rm sample}}})$,
or added with the probability $y^{\alpha_{{\rm sample}}}$ to the stack.
Depending on the choice of the sampling parameter 
($0\leq\alpha_{{\rm sample}}\leq1$)
either all the secondaries are kept in the cascade ($\alpha_{{\rm sample}}=0$)
or only some representative ones are retained. In particular, one secondary 
per interaction is retained on average for the default 
value $\alpha_{{\rm sample}}=1$. As compensation,
each particle in the cascade acquires a weight $w$ which
is augmented after each interaction as $w\to w/y^{\alpha_{{\rm sample}}}$.
The particles in the stack are ordered according to their energies.
After the interaction, the lowest energy particle is extracted
from the stack 
and traced further in the cascade process.

The optimal value of $\alpha_{\rm sample}$ depends on the typical energy of the
injected photons. If the latter is so low that the cascades consists on 
average of only few steps, reducing $\alpha_{\rm sample}$ may be advantageous
because the fluctuations are thereby reduced.

\subsection{Synchrotron losses}

Synchrotron energy losses of electrons are accounted for in the continuous
energy loss approximation using the interpolation formula \cite{baier},
\be \label{synch}
 \frac{dE}{dx} \ap
 \frac{m_{e}^{2}\,\chi^{2}}{[1+4.8(1+\chi)\,\ln(1+1.7\chi)+3.44\chi^{2}]^{2/3}}\,,
\ee
with $\chi=(p_\perp/m_e) (B/B_{\rm cr})$, where $p_\perp$ denotes
the momentum perpendicular to the magnetic field and 
$B_{\rm cr}=4.14\times 10^{13}$\,G the critical magnetic field.

\subsection{Angular deflection and time delay}

For the energies considered, $E_\gamma\gsim $\,MeV and $E_e\gsim 10$\,GeV, 
secondary particles
are emitted in the forward direction. Thus the angular deflection of the 
cascade particles results from the deflections of electrons in the
extragalactic magnetic field (EGMF). If the coherence scale of the
EGMF is much larger than electron mean free path $\lambda_{e}=R_e^{-1}$, an
elementary deflection angle of $i$-th electron in the cascade chain
can be calculated assuming the field to be regular over the distance
$d_{i}$ travelled by the electron,
\begin{equation}
\beta_{i}\simeq 0.52^\circ
\left(\frac{p_{\perp}}{{\rm TeV}}\right)^{-1}\left(\frac{d_{i}}{10\,{\rm kpc}}\right)\left(\frac{B}{10^{-15}{\rm G}}\right) \,,
\label{eq:beta-i}
\end{equation}
with $p_{\perp}$ being the momentum component perpendicular to the
local direction of the magnetic field. 
Inside the patch $j$ of a chosen coherence length, the deflection angles 
$\beta_{i}$ per electron path are summed up coherently,
$\beta_{j} =\sum_{i}\beta_{i}$. The deflections angles $\beta_{i}$ per 
coherent magnetic field patch are then summed quadratically in the 
random-walk approximation to obtain as total deflection, i.e.\
the angle $\beta$ between the initial and final photons in
the cascade, 
\begin{equation}
\beta=\sqrt{\sum_{i}\beta_{i}^{2}},
\label{eq:beta}
\end{equation}
In the small-angle approximation and assuming spherical symmetry,
the angle $\beta$ between the initial and 
final photons in the cascade is related to the emission angle $\alpha$ and 
the observation angle $\theta$ as $\alpha=\beta-\theta$, 
cf.\ Ref.~\cite{halo3}.

As the energy of the cascade particles quickly degrades along
the cascade chain, the largest contribution to $\beta$ comes from
the last electron in the chain. This allows us to approximate the
corresponding geometry by a triangular configuration 
and to obtain as relation between $\beta$ and $\theta$
\begin{equation}
 \sin\theta=\frac{x}{L}\,\sin\beta\,.
 \label{eq:theta}
\end{equation}
Here $x$ refers to the distance from the source S to the point P
where the final photon in the cascade branch has been created and
$L$ is the total distance between the source and the observer O.
For small $\theta$, we thus have
\begin{equation}
\theta=\frac{x}{L}\,\sin\beta\,.\label{eq:theta-small}
\end{equation}

The time delay $\Delta t_{\rm geo}$ of photons with respect to the
straight line propagation from the source is then
\begin{equation}
\Delta t_{\rm geo} \simeq x\,(1+\sin\alpha/\sin\theta)-L
         \simeq 2x\,(1-x/L)\,\sin^{2}\beta \,.
\label{eq:dt-fin}
\end{equation}
We add to this geometrical time delay $\Delta t_{\rm geo}$ the kinematical
time delay $\Delta t_{\rm kin}$ due to velocity $v<c$ of the electron, although
the latter is usually negligible.

\subsection{Cosmology}
The connection between redshift $z$, comoving distance $r$ and light-travel
time $t$ calculated for a flat Friedmann-Robertson-Walker universe
with $\Omega_\Lambda=0.7$ and  $\Omega_m=0.3$ is contained in the file
{\tt redshift}.

\section{Programme structure}
The programme is distributed among the files {\tt modules101.f90}, 
{\tt user101.f90}, {\tt init101.f90}, {\tt elmag101.f90} and {\tt aux101.f90}. 
The file {\tt modules101.f90} contains the definition of internal variables,
mathematical and physical constants; for standard applications of the 
programme no changes by the user are needed. 
The file {\tt user101.f90} contains the input/output subroutines 
developed by the user
for the desired task. An example file is discussed in Sec.~5. 
Data files of the used EBL backgrounds and the cosmological evolution
of the universe are provided in the directory {\tt Tables}. They are
read by the subroutines {\tt init\_EBL(myid)}, {\tt init\_arrays(myid)} and
the function {\tt aintIR(E,z)} inside the file {\tt init101.f90}.
Then  the function {\tt w\_EBL\_density\_tab(emin,zz)}
tabulates the weighted background photon density $I_\gamma$ defined in 
Eq.~(\ref{eq:I-gamma}), followed by the  tabulation of
the interaction rate in the the subroutine {\tt rate\_EBL\_tab(e0,zz,icq)}
and of the electron energy losses due to the emission of photons
with energy below the threshold in  {\tt eloss\_thr\_tab(e0,zz,icq)}.

We discuss now in more detail the subroutines and functions of the file 
{\tt elmag101.f90} which constitute the core of the programme:
\begin{itemize}
\item 
{\tt subroutine cascade(icq,e00,weight0,z\_in)}\\
Follows the evolution of the cascade initiated by a photon (${\tt icq}=0$) 
or an electron/positron (${\tt icq}=\pm 1$) injected at redshift {\tt z\_in} 
with energy {\tt e00} and weight {\tt weight0} until all secondary particles
have energies below the energy threshold {\tt ethr} or reached the observer 
at ${\tt z}=0$.
\item
{\tt subroutine angle\_delay(the2,xx,rcmb,theta,dt)}
\\
Determines the photon time-delay {\tt dt} and the observation
 angle {\tt theta} 
from the rms cascade deflection angle {\tt the2} and the photon emission
point {\tt xx}  by  the parent electron/positron.
\item
{\tt subroutine interaction(e0,x0,zz,t,weight,the1,the2,xxc,xx,dt,icq)}
\\
Handles one interaction with background photons: determines the c.m.\ energy 
{\tt sgam} of the reaction  via a call to {\tt sample\_photon} or 
{\tt sample\_electron(e0,zz,sgam,ierr)}, the energy fraction {\tt z} of
secondaries via a call to the functions {\tt zpair(sgam)} or 
{\tt zics(e0,sgam)},
and stores then the secondaries calling the subroutine {\tt store\_particle}.
\item 
{\tt subroutine store\_particle(e0,x0,zz,t,ze,weight,the1,the2,xxc,xx,dt,icq)}
\\
Decides if a produced secondary 
is stored using weighted sampling; if yes, 
it adds the secondary to the array {\tt event} and re-orders the array 
according to the particle energies.
\item 
{\tt subroutine get\_particle(e0,x0,zz,t,weight,the1,the2,xxc,xx,dt,icq)}
\\
Reads the secondary with the lowest energy out of the array {\tt event} and 
reduces the particle counter {\tt jcmb} by one.
\item
{\tt subroutine sample\_photon(e0,zz,sgam,ierr)} and
 {\tt sample\_electron(e0,zz,sgam,ierr)}
\\
Determines the cms energy {\tt sgam} of an interaction at redshift {\tt zz}.
\item 
{\tt double precision function w\_EBL\_density(emin,zz)}
\\
Determines the weighted background photon density $I_\gamma$ defined in
Eq. (\ref{eq:I-gamma}).  
\item 
{\tt function int\_point(e0,x0,zz,icq)}
\\
Finds the next interaction point for interaction with EBL photons.
\item
{\tt function sigpair(sgam)} and {\tt sigics(e0,sgam)}
\\
Calculate the pair production and inverse Compton cross section, respectively.
\item 
{\tt function zpair(sgam)} and {\tt zics(e0,sgam)}
\\
Determine the energy distribution in pair production and inverse 
Compton scattering, respectively.
\item 
{\tt function zsigics(e0,sgam)} 
\\
Calculates the electron energy losses per unit distance due to
photon emission below the chosen threshold
in  Compton scattering.
\item 
{\tt function zloss(e0,zz)}
\\
Interpolates the integrated energy loss due to emission of photons
below the threshold.
\item 
{\tt subroutine rate\_EBL(e0,zz,icq)}
\\
Interpolates the interaction rates $R_i$ on EBL photons.
\item {\tt function eloss\_syn(E,begmf)}
\\
Calculates the synchrotron losses according to Eq.~(\ref{synch}).
\item {\tt function themf(e0,dx,begmf)}
\\
Determines the deflection angle in the EGMF.
\end{itemize}

The file {\tt aux101.f90} contains auxiliary functions, e.g.\ the random
number generator {\tt psran} from Ref.~\cite{NR}.

\section{Example input and output}

\begin{figure}
\vskip-2.2cm
\includegraphics[width=0.45\linewidth,angle=0]{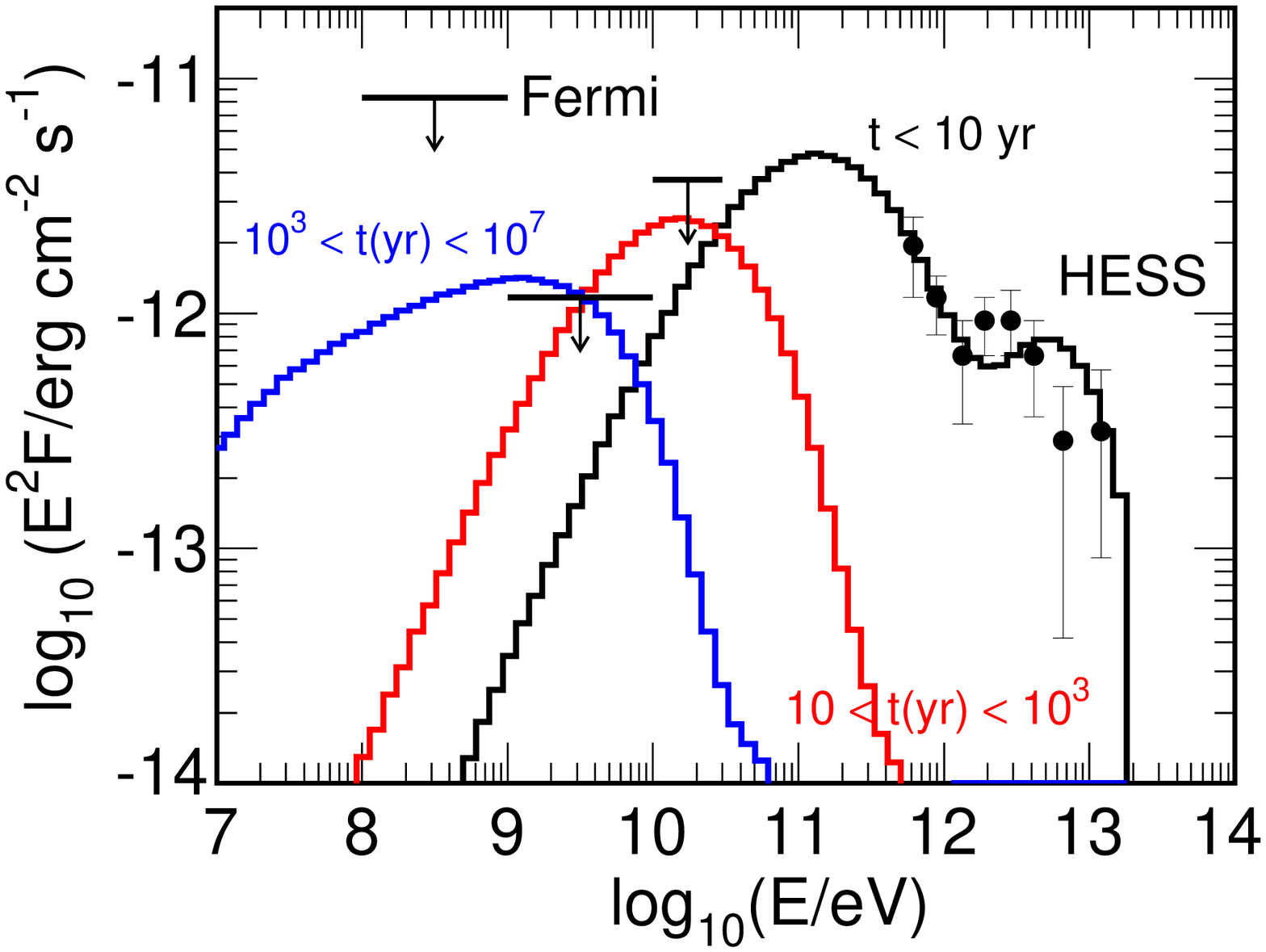}
\includegraphics[width=0.45\linewidth,angle=0]{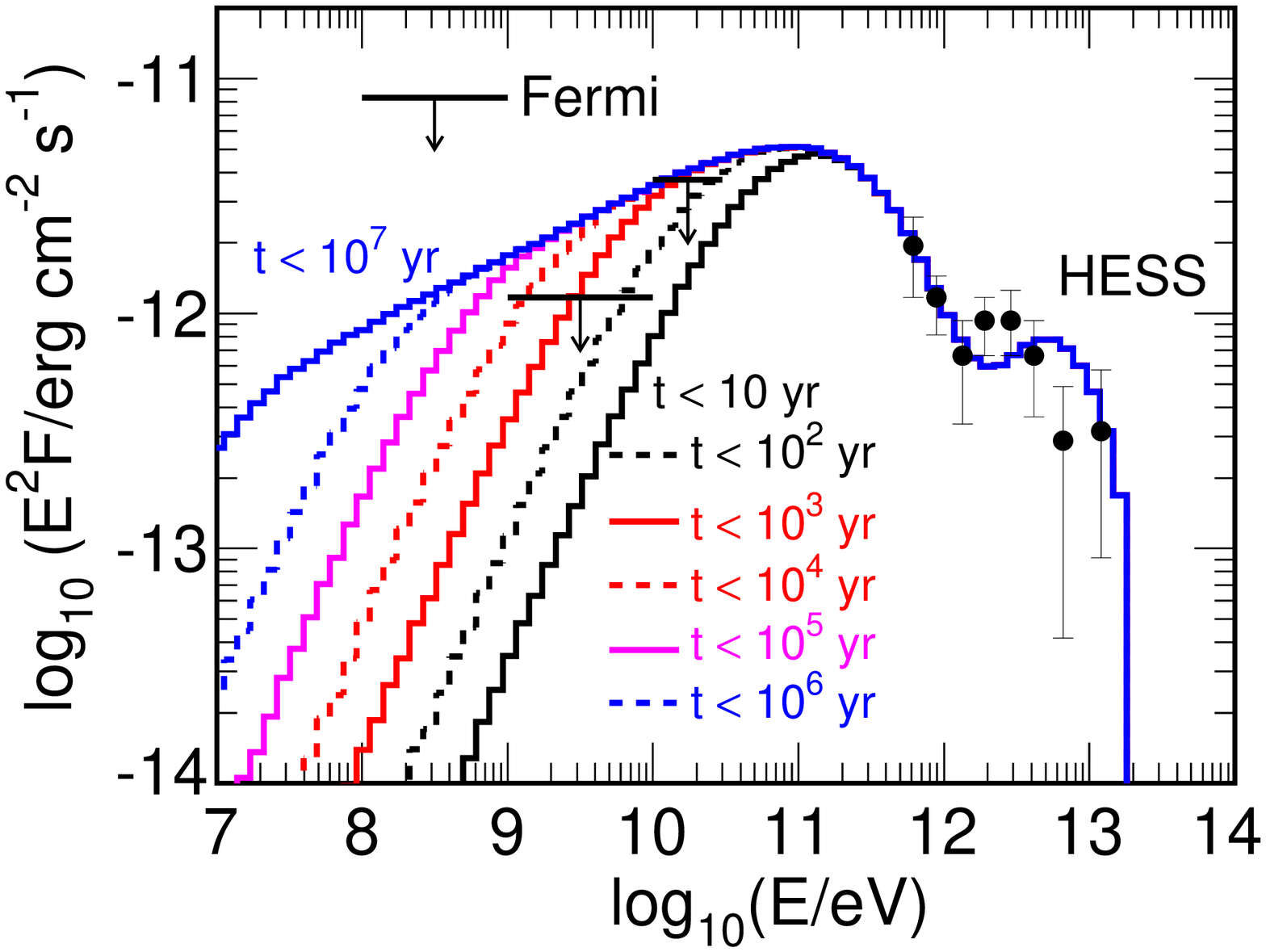}
\caption{
Fluence contained inside the 95\% confidence contour of the 
PSF of  Fermi-LAT as function of the time-delay for $B=10^{-17}$\,G;
left for individual time layers, right cumulative times.
\label{fig:time}} 
\end{figure}

The file {\tt user101.f90} is an example file for the input/output subroutines 
which should be developed by the user for the desired task. We discuss now 
the example contained in the distribution.

\subsection{Example input}

The input variables specified in the {\tt module user\_variables}
are: the choice of the EBL model ({\tt model}), the number of injected 
particles ({\tt nmax}), the jet opening angle of the source 
in degrees ({\tt th\_jet}), the sampling parameter  {\tt a\_smp}, 
the energy threshold {\tt ethr} for Compton scattering,
and the maximal photon energy {\tt egmax}. The last two parameters
serve also as minimal and maximal energy in the energy spectra produced
as output.
In {\tt subroutine user\_main(myid,nmax)} the initial redshift {\tt z}
and the particle type {\tt icq} of the injected particles is fixed.

\begin{verbatim}
  z = 0.14d0                                          ! initial redshift 
  do nl=1,nmax
     call initial_particle(e0,weight)                 ! generate initial energy
     icq = 0                                          ! (0 - gamma, +-1 - e+-)
     call cascade(icq,e0,weight,z)                    ! starts e/m cascade
  enddo
\end{verbatim}

The subroutine {\tt initial\_particle(e0,weight)} chooses the energy and
the weight of one initial particle in the energy range [{\tt emin,egmax}]
according to a broken power-law with exponents {\tt gam1} below {\tt ebreak}, 
and {\tt gam2} above.

The magnetic field $B$ is modeled as patches of uniform field-strength 
$|\mathbf B|$ of size $l_{\rm coh}$.
The value of the coherence length is fixed by the parameter
{\tt cohlnth}  in {\tt module user\_variables},  the  field-strength 
perpendicular to the propagation direction
in the {\tt function bemf(r)}.

\subsection{Example output}

The energy {\tt e0}, the observation angle {\tt theta} and the time delay 
{\tt dt} of secondary cascade particles with weight {\tt weight} of 
type {\tt icq} reaching the observer at $z=0$ are recorded by the
subroutine {\tt register(e0,theta,dt,weight,icq)} and binned
in various data arrays defined in the {\tt module user\_result}. 
All data arrays exists in two versions, e.g.\  {\tt spec(n\_bin,0:1)} 
and {\tt spec\_tot(n\_bin,0:1)}. Using MPI~\cite{MPI}, 
the former arrays contain the
result of a single process, which are summed by 
{\tt call MPI\_REDUCE} into {\tt spec\_tot(n\_bin,0:1)},
\begin{verbatim}
  n_array = 2*n_bin 
  call MPI_REDUCE(spec,spec_tot,n_array,MPI_DOUBLE_PRECISION,MPI_SUM,0, & \\
                  MPI_COMM_WORLD,ierr)        ! sum individal arrays spec
\end{verbatim}
Finally, the {\tt subroutine\ user\_output(n\_max,n\_proc)} writes the 
data arrays with the results in the files contained in the subdirectory
{\tt Data}.

The file {\tt spec\_diff} contains the normalised (diffuse) energy spectra 
of photons and electrons, in the format $E/eV$, $E^2dN_\gamma/dE$ and 
$E^2dN_e/dE$.
The file {\tt spec\_95} includes the energy spectra of photons inside 
and outside the 95\% area of the  point-spread function of Fermi-LAT,
in the format  $E/eV$, $E^2dN_\gamma/dE(\theta<\theta_{95})$ and 
$E^2dN_\gamma/dE(\theta<\theta_{95})$.  An approximation to the point-spread 
function of Fermi-LAT is defined in the {\tt function thereg\_en(en)}.

The right panel of Fig.~\ref{fig:lint} shows the energy spectra of photons 
arriving within the 95\% area of the point-spread function of Fermi-LAT
for different filling factors of the EGMF which can be chosen by
the parameter {\tt frac} in the function {\tt bemf}. Otherwise the default
values contained in the distributed file {\tt user101.f90} are used.
The photon fluence 
is compared to H.E.S.S. data~\cite{hess} and upper limits from 
Fermi-LAT~\cite{abdo09} for the TeV blazar 1ES~0229+200.

The files  {\tt spec\_95\_t} and {\tt spec\_95\_c} contain the energy 
spectra of photons arriving within the 95\% area of the point-spread 
function, with the time-delay binned in seven time intervals, 
$t<10\,{\rm yr}$,
$10\,{\rm yr}<10^2\,{\rm yr}$,\ldots $t>10^6\,{\rm yr}$.
The file {\tt spec\_95\_c} is the cumulative version of 
the distribution in  {\tt spec\_95\_t}.
Figure~\ref{fig:time} shows the fluence  inside the 95\% PSF of Fermi-LAT
for an injection spectrum $dN_\gamma/dE\propto E^{-2/3}$ at redshift $z=0.14$
with maximal energy $E_{\max} = 20$\,TeV.

\section{Possible extensions}

We discuss four possible applications of the simulation {\tt ELMAG} and the 
required extensions to perform them, ordered by the complexity of the
necessary changes and additions.
\paragraph{EGRB from dark matter decays or annihilations}

High-energy electrons and photons can be generated by decays or annihilations 
of sufficiently heavy dark matter particles. For instance, the annihilation 
mode $XX\to e^+e^-$ of the dark matter particle $X$ with mass $m_X$ would 
correspond to the injection of two electrons with energy $E_e=m_X$. 
The only necessary
addition for the calculation of the resulting EGRB is a subroutine choosing
the injection point according to the so-called boost factor $B(z)$ which 
accounts for the redshift dependent clustering of dark matter in galaxies.
Additionally, the desired fragmentation functions $dN_\gamma/dE$ 
of the $X$ particles should be 
included into the subroutine {\tt initial\_particle} in the case of
photons from hadronic decay or annihilation modes.

\paragraph{EGRB from UHECRs and cosmogenic neutrinos}

The Greisen-Zatsepin-Kuzmin cutoff is a steepening of the proton spectrum at 
the energy $E_{\rm GZK} \approx (4 - 5)\times 10^{19}$~eV, caused by photo-pion
production on the CMB. An additional signature for the presence of 
extragalactic 
protons in the cosmic ray flux and their interaction with CMB photons is the 
existence of ultrahigh energy cosmogenic neutrinos produced 
by charged pion decays~\cite{BZ}, while the corresponding flux of
cosmogenic neutrinos from  ultrahigh energy nuclei is suppressed. 
Photons and electrons from pion decay and 
$p + \gamma_{\rm CMB} \to p + e^+ + e^-$ pair-production lead to a 
contribution to the EGRB which can be used to limit  cosmic rays (CR) models and
fluxes of cosmogenic neutrinos. To perform this task, {\tt ELMAG}
has to be coupled to a program performing the propagation of  ultrahigh 
energy cosmic rays which provides secondary electrons and photons from
CR interactions as input. As the communication between the two program parts
is restricted to calls of the subroutine {\tt cascade(icq,e0,weight,z)},
such a combination should be straightforward. For an example where {\tt ELMAG}
was used in this context see Ref.~\cite{Berezinsky:2010xa}.

\paragraph{Extension to 3-dimensional cascades}

Going beyond the small-angle approximation requires to calculate the
actual trajectory of electrons solving the Lorentz equation. Additionally,
scalar quantities like {\tt x,xxc, e0, begmf,}\ldots
have to be changed into three-dimensional vectors and the {\tt type one\_event}
in the module {\tt stack} has to be adjusted. Finally, the image of
a three -dimensional cascade can be calculated using the method described
in Ref.~\cite{El09}. For an illustration of possible applications of 
{\tt ELMAG} to this problem see Ref.~\cite{Neronov:2010bd}.

\paragraph{Treatment of  interactions in sources}
Photons and electrons are generated often in sources containing
dense photon fields, as e.g.\ near the cores of active galactic nuclei.
In this case, electromagnetic cascades take place on non-thermal,
anisotropic photon backgrounds inside the source before the escaping
particles cascade on the EBL. In order to describe both cascades on
the EBL and inside the source, subroutines as e.g.\ 
{\tt init\_EBL} have to be doubled, adding a corresponding subroutine  
{\tt init\_source} for the photon field inside the source. All existing
subroutines which depend on the chosen EBL (i.e.\ {\tt use EBL\_fit})
have to adapted. In particular, the rejection mechanism in subroutines
as e.g.\  {\tt sample\_photon} has to adjusted. As a result, such an
extension requires considerable work and thorough tests of the changed code.
A short discussion 
of anisotropic photon fields is given in Ref.~\cite{Kachelriess:2008qx}.

\section{Summary}

We presented a Monte Carlo program for the simulation of 
electromagnetic cascades initiated by high-energy photons and electrons 
interacting with the extragalactic background light. The program
uses weighted sampling of the cascade development and treats Thomson
scattering below a chosen threshold in the continuous energy loss
approximation in order to speed up computations. 

Possible applications are the study of TeV blazars and the influence of
the EGMF on their spectra or the calculation of the contribution 
from ultrahigh energy cosmic rays or dark matter to the diffuse 
extragalactic gamma-ray background. As an illustration for possible 
applications we presented results for deflections and time-delays 
relevant for the derivation of limits on the EGMF 
 studying the spectra 
of TeV blazars. Other possible applications include e.g.\ 
the calculation of the contribution from ultrahigh energy cosmic rays or 
dark matter annihilations to the diffuse extragalactic gamma-ray background.

\section*{Acknowledgements}
We are grateful to Venya Berezinsky for valuable discussions and to Andrew
Taylor for cross-checking some of our results. 
This work was partially supported by the program Romforskning of 
the Norwegian Research Council.






\bibliographystyle{elsarticle-num}
\bibliography{<your-bib-database>}







\end{document}